\documentclass[aps,nofootinbib,floatfix,showpacs]{revtex4}
\usepackage{epsf,amsfonts,amssymb,amsbsy}

\newcommand{\Rsub}{\rm\scriptscriptstyle}
\begin{document}
\title{Decays of the $\boldsymbol B_{\boldsymbol c}$ meson}
\author{V.V.Kiselev}
\email{kiselev@th1.ihep.su}
\affiliation{Russian State Research Center "Institute for High
Energy
Physics",
Protvino, Moscow Region, 142281, Russia\\ Fax: +7-0967-744937}
\begin{abstract}
We summarize theoretical predictions on the decays of $B_c$ meson.
\end{abstract}

\maketitle

\hbadness=1500
\section{Introduction}
The investigation of long-lived heavy quarkonium $B_c$ composed of
quarks with different flavors can produce a significant progress
in the study of heavy quark dynamics, since the variation of bound
state conditions for the heavy quarks in various systems such as
the heavy-light hadrons or doubly heavy mesons and
baryons\footnote{See the review on the physics of baryons
containing two heavy quarks in \cite{QQq}.} provides us with the
different conditions in both the binding of quarks by the strong
interactions and the electroweak decays. In addition to the rich
fields of study such as the spectroscopy, production mechanism and
lifetime there is a possibility to get the model-independent
information on the CP-violating parameters in the heavy quark
sector\footnote{The extraction of angle $\gamma$ in the unitary
triangle derived from the charged current mixing in the heavy
quark sector can be obtained from the decays of doubly heavy
baryons, too, in the same manner \cite{CPQQq}.}
\cite{CPBc,CPBcKis}.

The first experimental observation of the $B_c$ meson by the CDF
collaboration \cite{cdf} confirmed the theoretical predictions on
its mass, production rate and lifetime
\cite{revbc,OPEBc,KT,KLO,KKL,exBc}. So, we could expect, that an
essential increase of statistics in the nearest future will
provide us with a new battle field in the study of long-lived
doubly heavy hadrons at Tevatron \cite{BRunII} and LHC
\cite{LHCB}.

Indeed, various hadronic matrix elements enter in the description
of weak decays. So, measuring the lifetimes and branching ratios
implies the investigation of quark confinement by the strong
interactions, which is important in the evaluation of pure quark
characteristics: masses and mixing angles in the CKM matrix, all
of which enter as constraints on the physics beyond the Standard
Model. More collection of hadrons with heavy quarks provides more
accuracy and confidence in the understanding of QCD dynamics and
isolation of bare quark values. So, a new lab for such
investigations is a doubly heavy long-lived quarkonium $B_c$.

Decays of the $B_c$ meson, containing two heavy quarks of
different flavors, were considered in the pioneering paper written
by Bjorken in 1986 \cite{Bj}. The report was devoted to the common
view onto the decays of hadrons with heavy quarks: the mesons and
baryons with a single heavy quark, the $B_c$ meson, the baryons
with two and three heavy quarks. A lot of efforts was recently
directed to study the long-lived doubly heavy
hadrons\footnote{Reviews on the physics of $B_c$ meson and doubly
heavy baryons can be found in refs. \cite{revbc,QQq},
respectively.} on the basis of modern understanding of QCD
dynamics in the weak decays of heavy flavors in the framework of
today approaches\footnote{See the program on the heavy flavour
physics at Tevatron in \cite{BRunII}.}: the Operator Product
Expansion, sum rules of QCD \cite{QCDSR} and NRQCD \cite{NRQCD},
and potential models adjusted due to the data on the description
of hadrons with a single heavy quark. Surprisingly, the Bjorken's
estimates of total widths and various branching fractions are
close to what is evaluated in a more strict manner. At present we
are tending to study some subtle effects caused by the influence
of strong forces onto the weak decays of heavy quarks, which
determines our understanding a probable fine extraction of {\sf
CP}-violation in the heavy quark sector.

The special role of heavy quarks in QCD is caused by the small
ratio of two physical scales: the heavy quark mass $m_Q$ is much
greater than the energetic scale of the quark confinement
$\Lambda_{\mbox{\sc qcd}}$. This fact opens a way to develop two
powerful tools in the description of strong interactions with the
heavy quarks. The first tool is the perturbative calculations of
Wilson coefficients determined by the hard corrections with the
use of renormalization group improvements. The second instrument
is the Operator Product Expansion (OPE) related with the small
virtuality of heavy quark in the bound state, which reveals itself
in many faces such as the general expansion of operators in
inverse powers of heavy quark mass and the QCD sum rules. A
specific form of OPE is the application for the heavy quark
lagrangian itself, which results in the effective theories of
heavy quarks. The effective theory is constructed under the choice
of its leading term appropriate for the system under study. So,
the kinetic energy can be neglected in the heavy-light hadrons to
the leading order. The corresponding effective theory is called
HQET \cite{HQET}. In the doubly heavy mesons the kinetic energy is
of the same order as the potential one, while the velocity of the
heavy quark motion is small, and we deal with the nonrelativistic
QCD \cite{NRQCD} and its developments by taking into account the
static energy in the potential NRQCD in a general form (pNRQCD)
\cite{pNRQCD} or under the correlation of quarkonium size and the
time for the formation of bound system with the improved scheme of
expansion in the heavy quark velocity (vNRQCD) \cite{vNRQCD}.

The $B_c$ meson allows us to use such advantages like the
nonrelativistic motion of $\bar b$ and $c$ quarks similar to what
is know in the heavy quarkonia $\bar b b$ and $\bar c c$, and
suppression of light degrees of freedom: the quark-gluon sea is
small in the heavy quarkonia. These two physical conditions imply
two small expansion parameters for $B_c$:
\begin{itemize}
\item the relative velocity of quarks $v$, \item the ratio of
confinement scale to the heavy quark mass $\Lambda_{QCD}/m_Q$.
\end{itemize}

The perturbative QCD remains actual in the effective theories,
since it is necessary for the calculation of Wilson coefficients
determined by hard corrections. More definitely, the perturbative
calculations determine both the matching of Wilson coefficients in
the effective theory with the full QCD and the anomalous
dimensions resulting in the evolution of the coefficients with the
variation of normalization point. In this respect we have to
mention that the pNRQCD results on the static potential and the
mass-dependent terms in the heavy quark-antiquark energy were
confirmed by vNRQCD after appropriate limits and some
calculational corrections in vNRQCD. In addition, the pNRQCD is a
powerful tool in the studies of both the spectroscopy and the
heavy quarkonium decays \cite{pNRQCD2}.

In contrast to the Wilson coefficients, the hadronic matrix
elements of operators composed by the effective fields of
nonrelativistic heavy quarks cannot be evaluated in the
perturbative manner. So, one should use the nonperturbative
methods such as the QCD sum rules, OPE for inclusive estimates and
potential models.

The measured $B_c$ lifetime is equal to
$$
\tau[B_c] = 0.46^{+0.18}_{-0.16}\pm 0.03\; {\rm ps,}
$$
which is close to the value expected by Bjorken. The $B_c$ decays
were, at first, calculated in the PM
\cite{PMBc,PMK,PML,PML2,vary,chch,ivanov,ISGW2,narod,CdF}, wherein
the variation of techniques results in close estimates after the
adjustment on the semileptonic decays of $B$ mesons. The OPE
evaluation of inclusive decays gave the lifetime and widths
\cite{OPEBc}, which agree with PM, if one sums up the dominating
exclusive modes. That was quite unexpected, when the SR of QCD
resulted in the semileptonic $B_c$ widths \cite{QCDSRBc}, which
are one order of magnitude less than those of PM and OPE. The
reason was the valuable role of Coulomb corrections, that implies
the summation of $\alpha_s/v$ corrections significant in the heavy
quarkonia, i.e. in the $B_c$ \cite{PML,KT,KLO,KKL}. At present,
all of mentioned approaches give the close results for the
lifetime and decay modes of $B_c$ at similar sets of parameters.
Nevertheless, various dynamical questions remain open:
\begin{itemize}
\item What is the appropriate normalization point of non-leptonic
weak lagrangian in the $B_c$ decays, which basically determines
its lifetime? \item What are the values of masses for the charmed
and beauty quarks? \item What are the implications of NRQCD
symmetries for the form factors of $B_c$ decays and mode widths?
\item How consistent is our understanding of hadronic matrix
elements, characterizing the $B_c$ decays, with the data on the
other heavy hadrons?
\end{itemize}
In this paragraph we shortly review the $B_c$ decays by
summarizing the theoretical predictions in the different
frameworks and discuss how direct experimental measurements can
answer the questions above.

\section{$\boldsymbol B_{\boldsymbol c} $ lifetime and inclusive decay
rates\label{2}}

The $B_c$-meson decay processes can be subdivided into three
classes:

1) the $\bar b$-quark decay with the spectator $c$-quark,

2) the $c$-quark decay with the spectator $\bar b$-quark and

3) the annihilation channel $B_c^+\rightarrow l^+\nu_l (c\bar s,
u\bar s)$, where $l=e,\; \mu,\; \tau$.

In the $\bar b \to \bar c c\bar s$ decays one separates also the
\underline{Pauli interference} with the $c$-quark from the initial
state. In accordance with the given classification, the total
width is the sum over the partial widths
$$
\Gamma (B_c\rightarrow X)=\Gamma (b\rightarrow X) +\Gamma
(c\rightarrow X)+\Gamma \mbox{(ann.)}+\Gamma\mbox{(PI)}.
$$
For the annihilation channel the  $\Gamma\mbox{(ann.)}$ width can
be reliably estimated in the framework of inclusive approach,
where one takes the sum of the leptonic and quark decay modes with
account for the hard gluon corrections to the effective four-quark
interaction of weak currents. These corrections result in the
factor of $a_1=1.22\pm 0.04$. The width is expressed through the
leptonic constant of $f_{B_c}\approx 400$ MeV. This estimate of
the quark-contribution does not depend on a hadronization model,
since a large energy release of the order of the meson mass takes
place. From the following expression, one can see that the
contribution by light leptons and quarks can be neglected,
$$
\Gamma \mbox{(ann.)} =\sum_{i=\tau,c}\frac{G^2_F}{8\pi}
|V_{bc}|^2f^2_{B_c}M m^2_i (1-m^2_i/m^2_{Bc})^2\cdot C_i\;,
\label{d3}
$$
where $C_\tau = 1$ for the $\tau^+\nu_\tau$-channel and $C_c
=3|V_{cs}|^2a_1^2 $ for the $c\bar s$-channel.

As for the non-annihilation decays, in the approach of the {\sf
Operator Product Expansion} for the quark currents of weak decays
\cite{OPEBc}, one takes into account the $\alpha_s$-corrections to
the free quark decays and uses the quark-hadron duality for the
final states. Then one considers the matrix element for the
transition operator over the bound meson state. The latter allows
one also to take into account the effects caused by the motion and
virtuality of decaying quark inside the meson because of the
interaction with the spectator. In this way the $\bar b\to \bar c
c\bar s$ decay mode turns out to be suppressed almost completely
due to the Pauli interference with the charm quark from the
initial state. Besides, the $c$-quark decays with the spectator
$\bar b$-quark are essentially suppressed in comparison with the
free quark decays because of a large bound energy in the initial
state.
\begin{table}[th]
\caption{The branching ratios of the $B_c$ decay modes calculated
in the framework of inclusive OPE approach, by summing up the
exclusive modes in the potential model \cite{PMK,PML} and
according to the semi-inclusive estimates in the sum rules of QCD
and NRQCD \cite{KLO,KKL}. } \label{t5}
\begin{center}
\begin{tabular}{|l|c|c|c|}
\hline
$B_c$ decay mode & OPE, \%  & PM, \% & SR, \%\\
\hline
$\bar b\to \bar c l^+\nu_l$ & $3.9\pm 1.0$  & $3.7\pm 0.9$  & $2.9\pm 0.3$\\
$\bar b\to \bar c u\bar d$  & $16.2\pm 4.1$ & $16.7\pm 4.2$ & $13.1\pm 1.3$\\
$\sum \bar b\to \bar c$     & $25.0\pm 6.2$ & $25.0\pm 6.2$ & $19.6\pm 1.9$\\
$c\to s l^+\nu_l$           & $8.5\pm 2.1$  & $10.1\pm 2.5$ & $9.0\pm 0.9$\\
$c\to s u\bar d$            & $47.3\pm 11.8$& $45.4\pm 11.4$& $54.0\pm 5.4$\\
$\sum c\to s$               & $64.3\pm 16.1$& $65.6\pm 16.4$& $72.0\pm 7.2$\\
$B_c^+\to \tau^+\nu_\tau$   & $2.9\pm 0.7$  & $2.0\pm 0.5$  & $1.8\pm 0.2$\\
$B_c^+\to c\bar s$          & $7.2\pm 1.8$  & $7.2\pm 1.8$  & $6.6\pm 0.7$\\
\hline
\end{tabular}
\end{center}
\label{inc}
\end{table}

In the framework of {\sf exclusive} approach, it is necessary to
sum up widths of different decay modes calculated in the potential
models. While considering the semileptonic decays due to the $\bar
b \to \bar c l^+\nu_l$ and $c\to s l^+\nu_l$ transitions, one
finds that the hadronic final states are practically saturated by
the lightest bound $1S$-state in the $(\bar c c)$-system, i.e. by
the $\eta_c$ and $J/\psi$ particles, and the $1S$-states in the
$(\bar b s)$-system, i.e. $B_s$ and $B_s^*$, which can only enter
the accessible energetic gap.

Further, the $\bar b\to \bar c u\bar d$ channel, for example, can
be calculated through the given decay width of $\bar b \to \bar c
l^+\nu_l$ with account for the color factor and hard gluon
corrections to the four-quark interaction. It can be also obtained
as a sum over the widths of decays with the $(u\bar d)$-system
bound states.

The results of calculation for the total $B_c$ width in the
inclusive OPE and exclusive PM approaches give the values
consistent with each other, if one takes into account the most
significant uncertainty related to the choice of quark masses
(especially for the charm quark), so that finally, we have
\begin{equation}
\left.\tau[B_c^+]\right._{\mbox{\small\sc ope,\,pm}}= 0.55\pm
0.15\; \mbox{ps,}
\end{equation}
which agrees with the measured value of $B_c$ lifetime.

\begin{figure}[th]
\setlength{\unitlength}{1.mm}
\begin{center}
\begin{picture}(130,95)
\put(0,5){\epsfxsize=120\unitlength \epsfbox{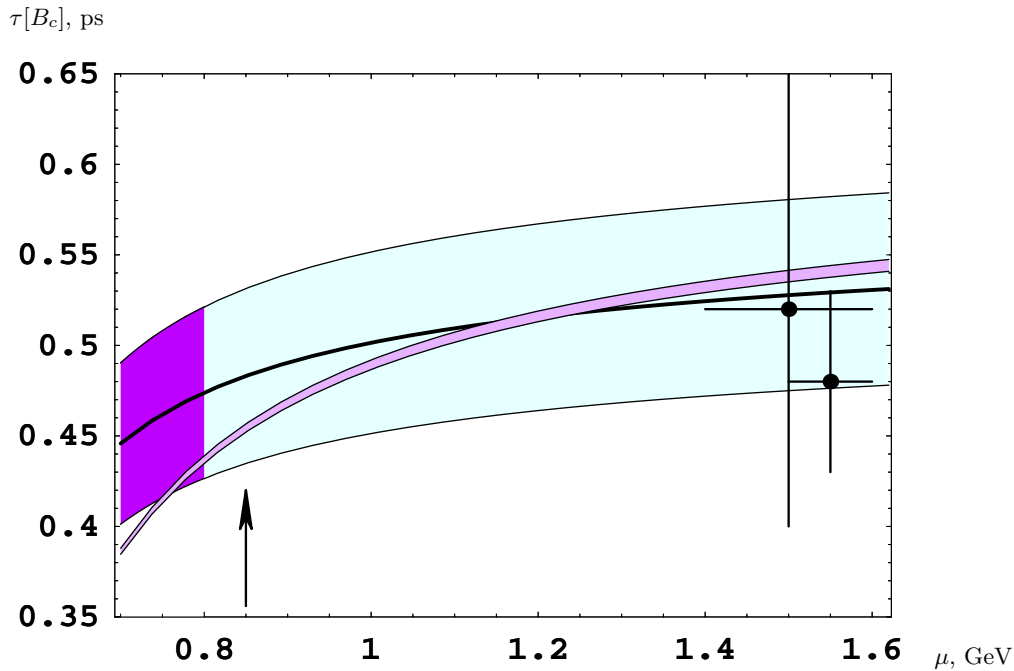}}
\put(0,92){$\tau[{B_c}] $, ps} \put(123,7){$\mu$, GeV}
\end{picture}
\end{center}
\caption{The $B_c$ lifetime calculated in QCD sum rules versus the
scale of hadronic weak lagrangian in the decays of charmed quark.
The wide shaded region shows the uncertainty of semi-inclusive
estimates, the dark shaded region is the preferable choice as
given by the lifetimes of charmed mesons. The dots represent the
values in OPE approach taken from ref. \cite{OPEBc}. The narrow
shaded region represents the result obtained by summing up the
exclusive channels with the variation of hadronic scale in the
decays of beauty anti-quark in the range of $1 <\mu_b < 5$ GeV.
The arrow points to the preferable prescription of $\mu =0.85$ GeV
as discussed in \cite{KKL}.} \label{life}
\end{figure}

The OPE estimates of inclusive decay rates agree with recent
semi-inclusive calculations in the sum rules of QCD and NRQCD
\cite{KLO,KKL}, where one assumed the saturation of hadronic final
states by the ground levels in the $c\bar c$ and $\bar b s$
systems as well as the factorization allowing one to relate the
semileptonic and hadronic decay modes. The coulomb-like
corrections in the heavy quarkonia states play an essential role
in the $B_c$ decays and allow one to remove the disagreement
between the estimates in sum rules and OPE. In contrast to OPE,
where the basic uncertainty is given by the variation of heavy
quark masses, these parameters are fixed by the two-point sum
rules for bottomonia and charmonia, so that the accuracy of SR
calculations for the total width of $B_c$ is determined by the
choice of scale $\mu$ for the hadronic weak lagrangian in decays
of charmed quark. We show this dependence in Fig. \ref{life},
where $\frac{m_c}{2} < \mu < m_c$ and the dark shaded region
corresponds to the scales preferred by data on the charmed meson
lifetimes.

Supposing the preferable choice of scale in the $c\to s$ decays of
$B_c$ to be equal to $\mu^2_{B_c} \approx (0.85\; {\rm GeV})^2$,
putting $a_1(\mu_{B_c}) =1.20$ and neglecting the contributions
caused by nonzero $a_2$ in the charmed quark decays \cite{KKL}, in
the framework of semi-inclusive sum-rule calculations we predict
\begin{equation}
\left.\tau[B_c]\right._{\mbox{\small\sc sr}} = 0.48\pm 0.05\;{\rm
ps},
\end{equation}
which agrees with the direct summation of exclusive channels
calculated in the next sections. In Fig. \ref{life} we show the
exclusive estimate of lifetime, too \cite{exBc}.

\section{Exclusive decays}

Predictions for the exclusive decays of $B_c$ are summarized in
Table \ref{common} at the fixed values of factors $a_{1,2}$ (see
below) and lifetime. We put the predicted values close to those of
QCD sum rules, which have an uncertainty about 15\% and basically
agree with the most of potential models, while marginal deviations
expected in some potential models are shown in square brackets.

In addition to the decay channels with the heavy charmonium
$J/\psi$ well detectable through its leptonic mode, one could
expect a significant information on the dynamics of $B_c$ decays
from the channels with a single heavy mesons, if an experimental
efficiency allows one to extract a signal from the cascade decays.
An interesting opportunity is presented by the relations for the
ratios in (\ref{fact}), which can shed light to characteristics of
the non-leptonic decays in the explicit form.

\begin{table*}[th]
\caption{Branching ratios of exclusive $B_c^+$ decays at the fixed
choice of factors: $a_1^c =1.20$ and $a_2^c=-0.317$ in the
non-leptonic decays of $c$ quark, and $a_1^b =1.14$ and
$a_2^b=-0.20$ in the non-leptonic decays of $\bar b$ quark. The
lifetime of $B_c$ is appropriately normalized by $\tau[B_c]
\approx 0.45$ ps. The numbers in square brackets present the
marginal values obtained in some potential models in order to show
possible range of variation.} \label{common}
\begin{center}
\begin{tabular}{|l|rr|}
\hline
~~~~~Mode & \multicolumn{2}{c|}{BR, \%}\\
\hline
 $B_c^+ \rightarrow \eta_c e^+ \nu$
 & 0.75 & [0.5]\\
 $B_c^+ \rightarrow \eta_c \tau^+ \nu$
 & 0.23 & [0.2]\\
 $B_c^+ \rightarrow \eta_c^\prime e^+ \nu$
 & 0.020 & [0.05]\\
 $B_c^+ \rightarrow \eta_c^\prime \tau^+ \nu$
 & 0.0016 & [-]\\
 $B_c^+ \rightarrow J/\psi e^+ \nu $
 & 1.9 & [1]\\
 $B_c^+ \rightarrow J/\psi \tau^+ \nu $
 & 0.48 & [0.35]\\
 $B_c^+ \rightarrow \psi^\prime e^+ \nu $
 & 0.094 & [0.2]\\
 $B_c^+ \rightarrow \psi^\prime \tau^+ \nu $
 & 0.008 & [-]\\
 $B_c^+ \rightarrow  D^0 e^+ \nu $
 & 0.004 & [0.02]\\
 $B_c^+ \rightarrow  D^0 \tau^+ \nu $
 & 0.002 & [0.08]\\
 $B_c^+ \rightarrow  D^{*0} e^+ \nu  $
 & 0.018  & [0.004]\\
 $B_c^+ \rightarrow  D^{*0} \tau^+ \nu  $
 & 0.008 & [0.016]\\
 $B_c^+ \rightarrow  B^0_s e^+ \nu  $
 & 4.03  & [1]\\
 $B_c^+ \rightarrow B_s^{*0} e^+ \nu  $
 & 5.06 & [1.2]\\
  $B_c^+ \rightarrow B^0 e^+ \nu  $
 & 0.34 & [0.08]\\
 $B_c^+ \rightarrow B^{*0} e^+ \nu  $
 & 0.58 & [0.15]\\
 $B_c^+ \rightarrow \eta_c \pi^+$
 & 0.20 & [0.12]\\
 $B_c^+ \rightarrow \eta_c \rho^+$
 & 0.42 & [0.3]\\
 $B_c^+ \rightarrow J/\psi \pi^+$
 & 0.13 & [0.08]\\
 $B_c^+ \rightarrow J/\psi \rho^+$
 & 0.40 & [0.2]\\
 $B_c^+ \rightarrow \eta_c K^+ $
 & 0.013 & [0.008]\\
 $B_c^+ \rightarrow \eta_c K^{*+}$
 & 0.020 & [0.018]\\
\hline
\end{tabular}
\begin{tabular}{|l|rr|}
\hline
~~~~~Mode & \multicolumn{2}{c|}{BR, \%}\\
\hline
 $B_c^+ \rightarrow J/\psi K^+$
 & 0.011 & [0.007]\\
 $B_c \rightarrow J/\psi K^{*+}$
 & 0.022 & [0.016]\\
 $B_c^+ \rightarrow D^+
\overline D^{\hspace{1pt}\raisebox{-1pt}{$\scriptscriptstyle 0$}}$
 & 0.0053 & [0.0018]\\
 $B_c^+ \rightarrow D^+
\overline D^{\hspace{1pt}\raisebox{-1pt}{$\scriptscriptstyle
*0$}}$
 & 0.0075 & [0.002]\\
 $B_c^+ \rightarrow  D^{\scriptscriptstyle *+}
\overline D^{\hspace{1pt}\raisebox{-1pt}{$\scriptscriptstyle 0$}}$
 & 0.0049 & [0.0009]\\
 $B_c^+ \rightarrow  D^{\scriptscriptstyle *+}
\overline D^{\hspace{1pt}\raisebox{-1pt}{$\scriptscriptstyle
*0$}}$
 & 0.033 & [0.003]\\
 $B_c^+ \rightarrow D_s^+ \overline
D^{\hspace{1pt}\raisebox{-1pt}{$\scriptscriptstyle 0$}}$
 & 0.00048 & [0.0001]\\
 $B_c^+ \rightarrow D_s^+
\overline D^{\hspace{1pt}\raisebox{-1pt}{$\scriptscriptstyle
*0$}}$
 & 0.00071 & [0.00012]\\
 $B_c^+ \rightarrow  D_s^{\scriptscriptstyle *+} \overline
D^{\hspace{1pt}\raisebox{-1pt}{$\scriptscriptstyle 0$}}$
 & 0.00045 & [0.00005]\\
 $B_c^+ \rightarrow  D_s^{\scriptscriptstyle *+}
\overline D^{\hspace{1pt}\raisebox{-1pt}{$\scriptscriptstyle
*0$}}$
 & 0.0026 & [0.0002]\\
 $B_c^+ \rightarrow \eta_c D_s^+$
 & 0.28 & [0.07]\\
 $B_c^+ \rightarrow \eta_c D_s^{*+}$
 & 0.27 & [0.07]\\
 $B_c^+ \rightarrow J/\psi D_s^+$
 & 0.17 & [0.05]\\
 $B_c^+ \rightarrow J/\psi D_s^{*+}$
 & 0.67 & [0.5]\\
 $B_c^+ \rightarrow \eta_c D^+$
 & 0.015 & [0.04]\\
 $B_c^+ \rightarrow \eta_c D^{*+}$
 & 0.010 & [0.002]\\
 $B_c^+ \rightarrow J/\psi D^+$
 & 0.009 & [0.002]\\
 $B_c^+ \rightarrow J/\psi D^{*+}$
 & 0.028 & [0.014]\\
 $B_c^+ \rightarrow B_s^0 \pi^+$
 & 16.4 & [1.6]\\
 $B_c^+ \rightarrow B_s^0 \rho^+$
 & 7.2 & [2.4]\\
 $B_c^+ \rightarrow B_s^{*0} \pi^+$
 & 6.5 & [1.3]\\
 $B_c^+ \rightarrow B_s^{*0} \rho^+$
 & 20.2 & [11]\\
\hline
\end{tabular}
\begin{tabular}{|l|rr|}
\hline
~~~~~Mode & \multicolumn{2}{c|}{BR, \%}\\
\hline
 $B_c^+ \rightarrow B_s^0 K^+$
 & 1.06 & [0.2]\\
 $B_c^+ \rightarrow B_s^{*0} K^+$
 & 0.37 & [0.13]\\
 $B_c^+ \rightarrow B_s^0 K^{*+}$
 & -- & \\
 $B_c^+ \rightarrow B_s^{*0} K^{*+}$
 & -- & \\
 $B_c^+ \rightarrow B^0 \pi^+$
 & 1.06 & [0.1]\\
 $B_c^+ \rightarrow B^0 \rho^+$
 & 0.96 & [0.2]\\
 $B_c^+ \rightarrow B^{*0} \pi^+$
 & 0.95 & [0.08]\\
 $B_c^+ \rightarrow B^{*0} \rho^+$
 & 2.57 & [0.6]\\
 $B_c^+ \rightarrow B^0 K^+$
 & 0.07 & [0.01]\\
 $B_c^+ \rightarrow B^0 K^{*+}$
 & 0.015 & [0.012]]\\
 $B_c^+ \rightarrow B^{*0} K^+$
 & 0.055 & [0.006]\\
 $B_c^+ \rightarrow B^{*0} K^{*+}$
 & 0.058 & [0.04]\\
 $B_c^+ \rightarrow B^+ \overline{K^0}$
 & 1.98 & [0.18]\\
 $B_c^+ \rightarrow B^+ \overline{K^{*0}}$
 & 0.43 & [0.09]\\
 $B_c^+ \rightarrow B^{*+} \overline{K^0}$
 & 1.60 & [0.06]\\
 $B_c^+ \rightarrow B^{*+} \overline{K^{*0}}$
 & 1.67 & [0.6]\\
 $B_c^+ \rightarrow B^+ \pi^0$
 & 0.037 & [0.004]\\
 $B_c^+ \rightarrow B^+ \rho^0$
 & 0.034 & [0.01]\\
 $B_c^+ \rightarrow B^{*+} \pi^0$
 & 0.033 & [0.003]\\
 $B_c^+ \rightarrow B^{*+} \rho^0$
 & 0.09 & [0.03]\\
 $B_c^+ \rightarrow \tau^+ \nu_\tau$
 & 1.6 & \\
 $B_c^+ \rightarrow c \bar s$
 & 4.9 & \\
\hline
\end{tabular}
\end{center}
\end{table*}

We have found that the $\bar b$ decay to the doubly charmed states
gives
$$
\mbox{Br}[B_c^+\to \bar c c\,c\bar s] \approx 1.39\%,
$$
so that in the absolute value of width it can be compared with the
estimate of spectator decay \cite{OPEBc},
\begin{eqnarray}
\left.\Gamma[B_c^+\to \bar c c\,c\bar s]\right|_{\mbox{\sc sr}}
&\approx & 20\cdot
10^{-15}\,\mbox{GeV},\nonumber \\[2mm]
\left.\Gamma[B_c^+\to \bar c c\,c\bar s]\right|_{\rm spect.}
&\approx & 90\cdot 10^{-15}\,\mbox{GeV}, \nonumber
\end{eqnarray}
and we find the suppression factor of about $1/4.5$. This result
is in agreement with the estimate in OPE \cite{OPEBc}, where a
strong dependence of negative term caused by the Pauli
interference on the normalization scale of non-leptonic weak
lagrangian was emphasized, so that at moderate scales one gets
approximately the same suppression factor, too.

To the moment we certainly state that the accurate direct
measurement of $B_c$ lifetime can provide us with the information
on both the masses of charmed and beauty quarks and the
normalization point of non-leptonic weak lagrangian in the $B_c$
decays (the $a_1$ and $a_2$ factors). The experimental study of
semileptonic decays and the extraction of ratios for the form
factors can test the spin symmetry derived in the NRQCD and HQET
approaches and decrease the theoretical uncertainties in the
corresponding theoretical evaluation of quark parameters as well
as the hadronic matrix elements, determined by the nonperturbative
effects caused by the quark confinement. The measurement of
branching fractions for the semileptonic and non-leptonic modes
and their ratios can inform on the values of factorization
parameters, which depend again on the normalization of
non-leptonic weak lagrangian. The charmed quark counting in the
$B_c$ decays is related to the overall contribution of $b$ quark
decays as well as with the suppression of $\bar b\to c\bar c \bar
s$ transition because of the destructive interference, which value
depends on the nonperturbative parameters (roughly estimated, the
leptonic constant) and non-leptonic weak lagrangian.

Thus, the progress in measuring the $B_c$ lifetime and decays
could enforce the theoretical understanding of what really happens
in the heavy quark decays at all.

\subsection{Semileptonic decays}
The semileptonic decay rates estimated in the QCD sum rules for
3-point correlators \cite{SR3pt} are underestimated in
\cite{QCDSRBc}, because large coulomb-like corrections were not
taken into account. The recent analysis of SR in \cite{KT,KLO,KKL}
decreased the uncertainty, so that the estimates agree with the
calculations in the potential
models.

\subsubsection{Coulomb corrections in the heavy quarkonia} 
\begin{figure}[th]
\setlength{\unitlength}{0.6mm}
\begin{center}
\begin{picture}(110,110)
\put(0, 0){\epsfxsize=110\unitlength \epsfbox{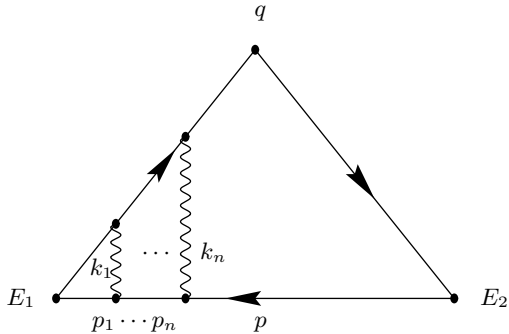}}
\put(0,40){$E_1$} \put(105,40){$E_2$} \put(55,35){$p$}
\put(55,104){$q$} \put(19,35){$p_1\cdots p_n$}
\put(30,50){$\cdots$} \put(18.5,46.5){$k_1$} \put(43,50){$k_n$}
\end{picture}
\end{center}
\normalsize

\vspace*{-2cm} \caption{The ladder diagram of the Coulomb-like
interaction.} \label{Coul-fig}
\end{figure}

For the
heavy quarkonium $\bar b c$, where the relative velocity of quark
movement is small, an essential role is taken by the Coulomb-like
$\alpha_s/v$-corrections. They are caused by the ladder diagram,
shown in Fig. \ref{Coul-fig}. It is well known that an account for
this corrections in two-point sum rules numerically leads to a
double-triple multiplication of Born value of spectral density
\cite{QCDSR}. In our case it leads to the finite renormalization
for $\rho_i$ \cite{KLO}, so that
\begin{equation}
 \rho^{c}_i={\cal C} \rho_i,
\label{ren}
\end{equation}
with
\begin{equation}
 {\cal C}^2=\left|\frac{\Psi^{\cal C}_{\bar b c}(0)}{\Psi^{\rm free}_{\bar b
 c}(0)}\right|^2=\frac{4\pi \alpha_s^{\cal
 C}}{3v}\,\frac{1}{\displaystyle 1-\exp\left(-\frac{4\pi\alpha_s^{\cal
 C}}{3v}\right)},
 \label{coul}
\end{equation}
where $v$ is the relative velocity of quarks in the $\bar b
c$-system,
\begin{equation}
v=\sqrt{1-\frac{4 m_b m_c}{p_1^2-(m_b-m_c)^2}},
\end{equation}
and the coupling constant of effective coulomb interactions
$\alpha_s^{\cal C}$ should be prescribed by the calculations of
leptonic constants for the appropriate heavy quarkonia as
described in the next section.

A similar coulomb factor appears in the vertex of heavy quarks
composing the final heavy quarkonium, in the case of $\bar c c$.

In order to fix such the parameters as the heavy quark masses and
effective couplings of coulomb exchange in the nonrelativistic
systems of heavy quarkonia with the same accuracy used in the
three-point sum rules, we explore the two-point sum rules of QCD
for the systems of $\bar c c$, $\bar b c$ and $\bar b b$. Thus, we
take into account the quark loop contribution with the coulomb
factors like that of (\ref{coul}). We keep this procedure despite
of current status of NRQCD sum rules for the heavy quarkonium,
wherein the three-loop corrections to the correlators are
available to the moment (see \cite{Hoang} for review), since for
the sake of consistency, the calculations should be performed in
the same order for both the three-point and two-point correlators.
This fact follows from the expression for the resonance term in
the three-point correlator, so that the form factor of transition
involves the normalization by the leptonic constants extracted
from the two-point sum rules. This procedure is taken since
calculations of two-loop corrections to the three-point
correlators are not available to the moment, unfortunately.

Then, the use of experimental values for the leptonic constants of
charmonium and bottomonium in addition to the consistent
description of spectral function moments in the two-point sum
rules allows us to extract the effective couplings of coulomb
exchange as well as the heavy quark masses in the heavy quarkonium
channels. We have found that the normalization of leptonic
constant is fixed by the appropriate choice of effective constant
for the coulomb exchange $\alpha_s^{\cal C}$, while the stability
is very sensitive to the prescribed value of heavy quark mass. The
physical meaning of such heavy quark masses is determined by the
threshold posing the energy at which the coulomb spectrum starts,
so that it is very close to the so-called `potential subtracted
masses' of heavy quarks, $m^{\rm PS}$ known in the literature (see
\cite{PS}) in the context of renormalon in the perturbative pole
mass \cite{renormalon}. We have found that the numerical values
obtained coincide with the appropriate $m^{\rm PS}$ within the
error bars.

\subsubsection{Primary modes}

In practice, the most constructive information is given by the
$\psi$ mode, since this charmonium is clearly detected in
experiments due to the pure leptonic decays \cite{cdf}. In
addition to the investigation of various form factors and their
dependence on the transfer squared, we would like to stress that
the measurement of decay to the excited state of charmonium, i.e.
$\psi^\prime$, could answer the question on the reliability of QCD
predictions for the decays to the excited states. We see that to
the moment the finite energy sum rules predict the width of
$B_c^+\to \psi^\prime l^+ \nu$ decays in a reasonable agreement
with the potential models if one takes into account an uncertainty
about 50\%.

\subsubsection{Relations between the form factors\label{4}}

In the limit of infinitely heavy quark mass, the NRQCD and HQET
lagrangians possess the spin symmetry, since the heavy quark spin
is decoupled in the leading approximation. The most familiar
implication of such the symmetry is the common Isgur-Wise function
determining the form factors in the semileptonic decays of singly
heavy hadrons.

In contrast to the weak decays with the light spectator quark, the
$B_c$ decays to both the charmonia $\psi$ and $\eta_c$ and
$B_s^{(*)}$ involve the heavy spectator, so that the spin symmetry
works only at the recoil momenta close to zero, where the
spectator enters the heavy hadron in the final state with no hard
gluon rescattering. Hence, in a strict consideration we expect the
relations between the form factors in the vicinity of zero recoil.
The normalization of common form factor is not fixed, as was in
decays of hadrons with a single heavy quark, since the heavy
quarkonia wave-functions are flavour-dependent. Nevertheless, in
practice, the ratios of form factors as fixed at a given zero
recoil point are broken only by the different dependence on the
transfer squared, that is not significant in real numerical
estimates in the restricted region of physical phase space.

As for the implications of spin symmetry for the form factors of
decay, in the soft limit for the transitions $B_c^+\to
\psi(\eta_c) e^+\nu$
\begin{eqnarray}
v_1^{\mu} &\neq & v_2^{\mu}, \label{cs}\\
w &=& v_1\cdot v_2\to 1,\nonumber
\end{eqnarray}
where $v_{1,2}^{\mu} = p_{1,2}^{\mu}/\sqrt{p_{1,2}^2}$ are the
four-velocities of heavy quarkonia in the initial and final
states, we derive the relations \cite{KLO}
\begin{widetext}
\begin{eqnarray}
f_{+}(c_1^{P}\cdot{\cal M}_2 - c_2^{P}{\cal M}_1) -
f_{-}(c_1^{P}\cdot{\cal M}_2 + c_2^{P}\cdot{\cal M}_1) = 0, &&
\quad F_{0}^{A}\cdot c_V - 2 c_{\epsilon}\cdot F_V{\cal M}_1{\cal
M}_2  = 0,
\label{Fsym}\\[2mm]
F_{0}^{A}(c_1 + c_2) - c_{\epsilon}{\cal M}_1 (F_{+}^{A}({\cal
M}_1 + {\cal M}_2) +  F_{-}^{A}({\cal M}_1 - {\cal M}_2))  = 0,
&&\quad F_{0}^{A}c_1^{P} + c_{\epsilon}\cdot{\cal M}_1(f_{+} +
f_{-})  = 0,
\end{eqnarray}
where
\begin{equation}
\begin{array}{lll}
c_{\epsilon} = -2,&~~ \displaystyle c_1 = -\frac{m_3(3m_1 +
m_3)}{4m_1m_2},&~~ \displaystyle c_2 = \frac{1}{4m_1m_2}(4m_1m_2 +
m_1m_3 + 2m_2m_3 +
m_3^2),\\[4mm]
\displaystyle c_1^{P} = 1 + \frac{m_3}{2m_1} -
\frac{m_3}{2m_2},&~~ \displaystyle c_2^{P} = 1 - \frac{m_3}{2m_1}
+ \frac{m_3}{2m_2},&~~ \displaystyle c_V =
-\frac{1}{2m_1m_2}(2m_1m_2 + m_1m_3 + m_2m_3),
\end{array}
\end{equation}
\end{widetext}
so that $m_1$ is the mass of decaying quark, $m_2$ is the quark
mass of decay product, and $m_3$ is the mass of spectator quark,
while ${\cal M}_1=m_1+m_2$, ${\cal M}_2=m_2+m_3$.

The SR estimates of form factors show a good agreement with the
relations, whereas the deviations can be basically caused by the
difference in the $q^2$-evolution of form factors from the zero
recoil point, that can be neglected within the accuracy of SR
method for the transitions of $B_c\to \bar c c$ as shown in
\cite{KLO}.

In the same limit for the semileptonic modes with a single heavy
quark in the final state we find that the ambiguity in the `light
quark
propagator' 
(strictly, we deal with the uncertainty in the spin structure of
amplitude
because of light degrees of freedom) 
restricts the number of relations, and we derive
\begin{widetext}
\begin{equation}
f_{+}(\bar c_1^{P}\cdot{\cal M}_2 - \bar c_2^{P}{\cal M}_1) -
f_{-}(\bar c_1^{P}\cdot{\cal M}_2 + \bar c_2^{P}\cdot{\cal M}_1) =
0,\quad F_{0}^{A}\cdot \bar c_V - 2 \bar c_{\epsilon}\cdot
F_V{\cal M}_1{\cal M}_2 = 0,\quad \label{Fsymq} F_{0}^{A}\bar
c_1^{P} + \bar c_{\epsilon}\cdot{\cal M}_1(f_{+} + f_{-}) = 0,
\end{equation}
where
\begin{equation}
\bar c_{\epsilon} = -2,\quad \bar c_V = -1-\tilde
B-\frac{m_3}{2m_1},\quad \bar c_1^{P} = 1-\tilde
B+\frac{m_3}{2m_1},\quad \bar c_2^{P} = 1+\tilde
B-\frac{m_3}{2m_1},
\end{equation}
\end{widetext}
so that $m_2$ is the mass of the light quark. The parameter
$\tilde B$ has the form
\begin{equation}
\tilde B=-\frac{2m_1+m_3}{2m_1}+\frac{4m_3(m_1+m_3)F_V}{F_0^A}.
\end{equation}
The $1/m_Q$-deviations from the symmetry relations in the decays
of $B_c^+\to B_s^{(*)} e^+\nu$ are about 10-15 \%, as found in the
QCD sum rules considered in \cite{KKL}.

Next, we investigate the validity of spin-symmetry relations in
the $B_c$ decays to $B^{(*)}$, $D^{(*)}$ and $D_s^{(*)}$. The
results of estimates for the $f_{\pm}$ evaluated by the symmetry
relations with the inputs given by the form factors $F_V$ and
$F_0^A$ extracted from the sum rules have been compared with the
values calculated in the framework of sum rules.

We have found that the uncertainty in the estimates is basically
determined by the variation of pole masses in the
$q^2$-dependencies of form factors, which govern the evolution
from the zero recoil point to the zero transfer squared. So, the
variation of $M_{\rm pole}[f_{\pm}]$ in the range of $4.8-5$ GeV
for the transitions of $B_c\to D^{(*)}$ and $B_c\to D_s^{(*)}$
results in the 30\%-uncertainty in the form factors. Analogously,
the variation of $M_{\rm pole}[f_{\pm}]$ in the range of $1.5-1.9$
GeV for the transition of $B_c\to B^{(*)}$ results in the
uncertainty about 35\%.

Note, that the combinations of relations given above reproduce the
only equality \cite{Jenk}, which was found for each mode in the
strict limit of $v_1=v_2$.

\subsection{Leptonic decays}

The dominant leptonic decay of $B_c$ is given by the $\tau
\nu_\tau$ mode (see Table \ref{inc}). However, it has a low
experimental efficiency of detection because of hadronic
background in the $\tau$ decays or a missing energy. Recently, in
refs. \cite{radlep} the enhancement of muon and electron channels
in the radiative modes was studied. The additional photon allows
one to remove the helicity suppression for the leptonic decay of
pseudoscalar particle, which leads, say, to the double increase of
muonic mode.

\subsubsection{Leptonic constant of $B_{c}$}

In the NRQCD approximation for the heavy quarks, the calculation of leptonic
constant for the heavy quarkonium with the two-loop accuracy requires the
matching of NRQCD currents with the currents in full QCD,
$$
J_\nu^{\Rsub QCD}= \bar Q_1 \gamma_5\gamma_\nu Q_2, \;\;\; {\cal
J}_\nu^{\Rsub NRQCD} = -\chi^\dagger  \phi \; v_\nu,
$$
where we have introduced the following notations: $Q_{1,2}$ are the
relativistic quark fields, $\chi$ and $\phi$ are the nonrelativistic spinors of
anti-quark and quark, $v$ is the four-velocity of heavy quarkonium, so that
\begin{equation}
J_\nu^{\Rsub QCD} = {\cal K}(\mu_{\rm hard}; \mu_{\rm fact})\cdot
{\cal J}_\nu^{\Rsub NRQCD}(\mu_{\rm fact}), \label{match}
\end{equation}
where the scale $\mu_{\rm hard}$ gives the normalization point for the matching
of NRQCD with full QCD, while $\mu_{\rm fact}$ denotes the normalization point
for the calculations in the perturbation theory of NRQCD.

For the pseudoscalar heavy quarkonium composed of heavy quarks
with the different flavors, the Wilson coefficient ${\cal K}$ is
calculated with the two-loop
accuracy  
\begin{eqnarray}
{\cal K}(\mu_{\rm hard}; \mu_{\rm fact}) &=& 1 + c_1\,
\frac{\alpha_s^{\overline{\Rsub MS}}(\mu_{\rm
hard})}{\pi}+
c_2(\mu_{\rm hard}; \mu_{\rm
fact})\left(\frac{\alpha_s^{\overline{\Rsub MS}}(\mu_{\rm
hard})}{\pi}\right)^2\hspace*{-4pt}, \label{kfact}
\end{eqnarray}
and $c_{1,2}$ are explicitly given in Refs. \cite{braflem} and
\cite{ov}, respectively. The anomalous dimension of factor ${\cal K}(\mu_{\rm
hard}; \mu_{\rm fact})$ in NRQCD is defined by
\begin{equation}
\frac{d \ln{\cal K}(\mu_{\rm hard}; \mu)}{d \ln \mu} = \sum_{k=1}^{\infty}
\gamma_{[k]} \left(\frac{\alpha_s^{\overline{\Rsub MS}}(\mu)}{4\pi}\right)^k,
\label{anom}
\end{equation}
whereas the two-loop calculations\footnote{We use ordinary notations for the
invariants of $SU(N_c)$ representations: $C_F=\frac{N_c^2-1}{2 N_c}$, $C_A=
N_c$, $T_F = \frac{1}{2}$, $n_f$ is a number of ``active'' light quark
flavors.} give
\begin{eqnarray}
\gamma_{[1]} & = & 0,\\
\gamma_{[2]} & = & -8 \pi^2 C_F
\left[\left(2-\frac{(1-r)^2}{(1+r)^2}\right) C_F + C_A\right],
\label{g2}
\end{eqnarray}
where $r$ denotes the ratio of heavy quark masses. The initial
condition for the evolution of factor ${\cal K}(\mu_{\rm hard};
\mu_{\rm fact})$ is given by the matching of NRQCD current with
full QCD at $\mu_{\rm fact} = \mu_{\rm hard}$.

The leptonic constant is defined in the following way:
\begin{equation}
\langle 0| J_\nu^{\Rsub QCD} |\bar Q Q \rangle = v_\nu f_{\bar Q Q
} M_{\bar Q Q }.
\end{equation}
In full QCD the axial vector current of quarks has zero anomalous
dimension, while in NRQCD the current ${\cal J}_\nu^{\Rsub NRQCD}$
has the nonzero anomalous dimension, so that in accordance with
(\ref{match})--(\ref{g2}), we find
\begin{equation}
\langle 0| {\cal J}_\nu^{\Rsub NRQCD}(\mu) |\bar Q Q \rangle =
{\cal A}(\mu)\; v_\nu f_{\bar Q Q }^{\Rsub NRQCD} M_{\bar Q Q },
\label{a}
\end{equation}
where, in terms of nonrelativistic quarks, the leptonic constant for the heavy
quarkonium is given by the well-known relation with the wave function at the
origin
\begin{equation}
f_{\bar Q Q}^{\Rsub NRQCD} = \sqrt{\frac{12}{M_{\bar Q Q}}}\;
|\Psi_{\bar Q Q}(0)|, \label{wave}
\end{equation}
and the value of wave function in the leading order is determined
by the solution of Schr\"odinger equation with the static
potential, so that we isolate the scale dependence of NRQCD
current in the factor ${\cal A(\mu)}$, while the leptonic constant
$f_{\bar Q Q}^{\Rsub NRQCD}$ is evaluated at a fixed normalization
point $\mu=\mu_0$, which will be attributed below. It is evident
that
\begin{equation}
f_{\bar QQ} = f_{\bar QQ}^{\Rsub NRQCD} {\cal A}(\mu_{\rm
fact})\cdot {\cal K}(\mu_{\rm hard}; \mu_{\rm fact}), \label{cc}
\end{equation}
and the anomalous dimension of ${\cal A}(\mu_{\rm fact})$ should compensate the
anomalous dimension of factor ${\cal K}(\mu_{\rm hard}; \mu_{\rm fact})$, so
that in two loops we have got
\begin{equation}
\frac{d \ln{\cal A}(\mu)}{d \ln \mu} = - \gamma_{[2]}
\left(\frac{\alpha_s^{\overline{\Rsub MS}}(\mu)}{4\pi}\right)^2.
\label{anoma}
\end{equation}
The physical meaning of ${\cal A}(\mu)$ is clearly determined by the relations
of (\ref{a}) and (\ref{cc}): this factor gives the normalization of matrix
element for the current of nonrelativistic quarks expressed in terms of wave
function for the two-particle quark state (in the leading order of inverse
heavy quark mass in NRQCD). Certainly, in this approach the current of
nonrelativistic quarks is factorized from the quark-gluon sea, which is a
necessary attribute of hadronic state, so that, in general, this physical state
can be only approximately represented as the two-quark bound state. In the
consideration of leptonic constants in the framework of NRQCD, this
approximation requires to introduce the normalization factor ${\cal A}(\mu)$
depending on the scale.

The renormalization group equation of (\ref{anoma}) is simply integrated out,
so that
\begin{equation}
{\cal A}(\mu) = {\cal A}(\mu_0)\; \left[ \frac{\beta_0+\beta_1
{\displaystyle\frac{\alpha_s^{\overline{\Rsub
MS}}(\mu)}{4\pi}}}{\beta_0+\beta_1 {\displaystyle
\frac{\alpha_s^{\overline{\Rsub MS}}(\mu_0)}{4\pi}}}
\right]^{\displaystyle\frac{\gamma_{[2]}}{2\beta_1}}, \label{RG2}
\end{equation}
where $\beta_0 = \frac{11}{3} C_A - \frac{4}{3} T_F n_f$, and
$\beta_1 = \frac{34}{3} C_A^2 - 4 C_F T_F n_f  - \frac{20}{3} C_A
T_F n_f$. A constant of integration could be defined so that at a
scale $\mu_0$ we would get ${\cal A}(\mu_0)=1$. Thus, in the
framework of NRQCD we have got the parametric dependence of
leptonic constant estimates on the scale $\mu_0$, which has the
following simple interpretation: the normalization of matrix
element for the current of nonrelativistic quarks at $\mu_0$ is
completely given by the wave function of two-quark bound state. At
other $\mu\ne \mu_0$ we have to introduce the factor ${\cal
A}(\mu)\ne 1$, so that the approximation of hadronic state by the
two-quark wavefunction becomes inexact.

Following the method described in \cite{KKO,KLPS}, we estimate the
wave function of $\bar b c$ quarkonium in the nonrelativistic
model with the static potential given by \cite{KKO}. Details of
calculations are presented in \cite{fBc-V}.

\begin{figure}[th]
\setlength{\unitlength}{0.96mm}
\begin{center}
\begin{picture}(100,110)
\put(-5,5){\epsfxsize=100\unitlength \epsfbox{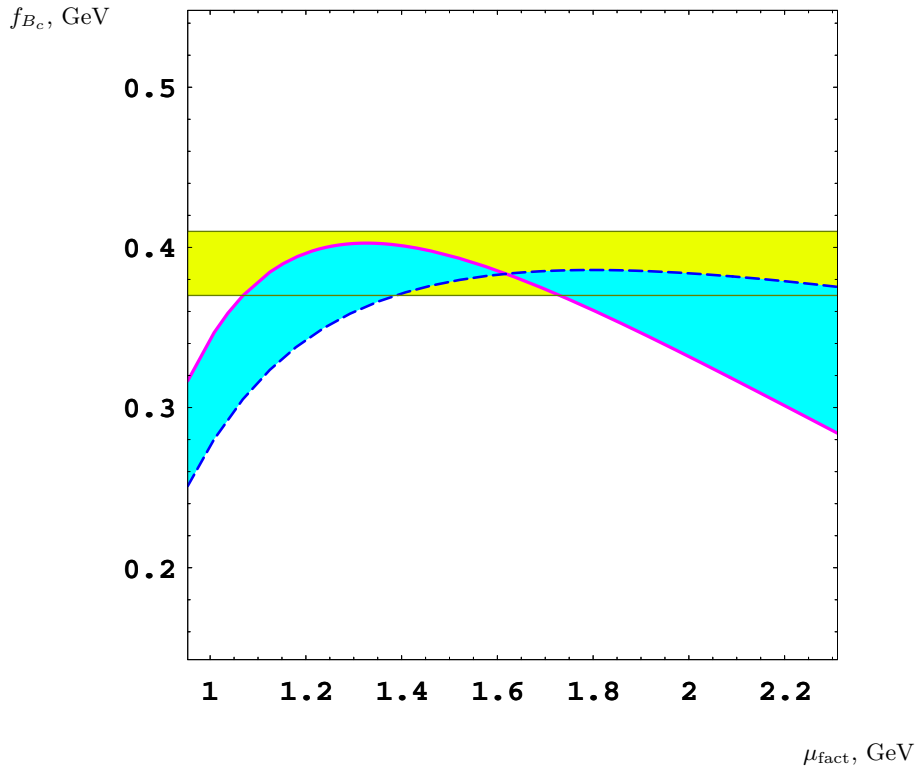}}
\put(90,0){$\mu_{\rm fact}$, GeV} \put(-20,102){$f_{B_c}$, GeV}
\end{picture}
\end{center}

\caption{The leptonic constant of ground pseudoscalar state in the
system of heavy quarkonium $\bar b c$ is presented versus the soft
scale of normalization. The shaded region restricted by curves
corresponds to the change of hard scale from $\mu_{\rm hard} = 3$
GeV (the dashed curve) to $\mu_{\rm hard} = 2$ GeV (the solid
curve) with the initial condition for the evolution of
normalization factor ${\cal A}(\mu_{\rm fact})$ posed in the form
of ${\cal A}(1.2\;{\rm GeV})=1$ and ${\cal A}(1.\;{\rm GeV})=1$,
respectively, in the matrix element of current given in the
nonrelativistic representation. The horizontal band is the limits
expected from the QCD sum rules \cite{sr} and scaling relations
for the leptonic constants of heavy quarkonia \cite{scale}. In the
cross-point, the leptonic constant of $B_c$ weakly depends on the
parameters given by the hard scale of matching as well as the
scale of the initial normalization.} \label{fB_c}
\end{figure}

The result of calculation for the leptonic constant of $B_c$ in
the potential approach is shown in Fig. \ref{fB_c}. We chose the
values of $\mu_{\rm hard}$ in a range, so that the stable value of
leptonic constant would be posed at the scale $\mu_{\rm fact}<
m_c^{\rm pole}$, which is the condition of consistency for the
NRQCD approach. Then one can observe the position of $\mu_{\rm
fact}\approx 1.5-1.6$ GeV, where the estimate of $f_{B_c}$ is
independent of the choice of $\mu_{\rm hard}$ and $\mu_0$, and the
maximal values at the different choices of $\mu_{\rm hard}$ are
also close to each other.

The final result of two-loop calculations is
\begin{equation}
f_{B_c} = 395\pm 15\; {\rm MeV}. \label{fin}
\end{equation}
It should be compared with the estimate of potential model itself
without the matching
\begin{equation}
f^{\Rsub NRQCD}_{B_c} = 493\; {\rm MeV},
\end{equation}
which indicates the magnitude of the correction about 20\%.
Furthermore, the calculations in the same potential model with the
one-loop matching \cite{KKO} gave
\begin{equation}
f^{1-loop}_{B_c} = 400\pm 45\; {\rm MeV},
\end{equation}
where the uncertainty is significantly greater than in the
two-loop procedure, since at the one-loop level we have no stable
point in the scale dependence of the result. Therefore, in
contrast to the discussion given in \cite{ov} we see that the
correction is not crucially large, but it is under control in the
system of $B_c$. The reason for such the claim on the reliability
of result is caused by two circumstances. First, the one-loop
anomalous dimension of NRQCD current is equal to zero. Therefore,
we start the summation of large logs in the framework of
renormalization group (RG) with the expressions in (\ref{anoma})
and (\ref{RG2}). Second, after such the summation of large logs
the three-loop corrections could be considered as small beyond the
leading RG logs.

The result on $f_{B_c}$ is in agreement with the scaling relation
derived from the quasi-local QCD sum rules \cite{scale}, which use
the regularity in the heavy quarkonium mass spectra, i.e. the fact
that the splitting between the quarkonium levels after the
averaging over the spins of heavy quarks weakly depends on the
quark flavors. So, the scaling law for the S-wave quarkonia has
the form
\begin{equation}
\frac{f_n^2}{M_n}\;\left(\frac{M_n}{M_1}\right)^2\;
\left(\frac{m_1+m_2}{4\mu_{12}}\right)^2
 = \frac{c}{n}, \label{2.3}
\end{equation}
where $n$ is the radial quantum number, $m_{1,2}$ are the masses
of heavy quarks composing the quarkonium, $\mu_{12}$ is the
reduced mass of quarks, and $c$ is a dimensional constant
independent of both the quark flavors and the level number $n$.
The value of $c$ is determined by the splitting between the $2S$
and $1S$ levels, or the average kinetic energy of heavy quarks,
which is independent of the quark flavors and $n$ with the
accuracy accepted. The accuracy depends on the heavy quark masses,
and it is discussed in \cite{scale} in detail. The parameter $c$
can be extracted from the known leptonic constants of $\psi$ and
$\Upsilon$, so that the scaling relation gives
$$
f_{B^*_c}\approx 400\;{\rm MeV}
$$
for the vector state. The difference between the leptonic
constants for the pseudoscalar and vector $1S$-states is caused by
the spin-dependent corrections, which are small. Numerically, we
get $|f_{B^*_c}-f_{B_c}|/f_{B^*_c}< 3\%$, hence, the estimates
obtained from the potential model and the scaling relation is in a
good agreement with each other.

\subsection{Non-leptonic modes\label{6}}

In comparison with the inclusive non-leptonic widths, which can be
estimated in the framework of quark-hadron duality (see Table
\ref{inc}), the calculations of exclusive modes usually involves
the approximation of factorization \cite{fact}, which, as
expected, can be quite accurate for the $B_c$, since the
quark-gluon sea is suppressed in the heavy quarkonium. Thus, the
important parameters are the factors $a_1$ and $a_2$ in the
non-leptonic weak lagrangian, which depend on the normalization
point suitable for the $B_c$ decays.

The agreement of QCD SR estimates for the non-leptonic decays of
charmed quark in $B_c$ with the values predicted by the potential
models is rather good for the direct transitions with no
permutation of color lines, i.e. the class I processes with the
factor of $a_1$ in the non-leptonic amplitude determined by the
effective lagrangian. In contrast, the sum rule predictions are
significantly enhanced in comparison with the values calculated in
the potential models for the transitions with the color
permutation, i.e. for the class II processes with the factor of
$a_2$ (see Table \ref{common}).

Further, for the transitions, wherein the interference is
significantly involved, the class III processes, we find that the
absolute values of different terms given by the squares of $a_1$
and $a_2$ calculated in the sum rules are in agreement with the
estimates of potential models. Taking into account the negative
value of $a_2$ with respect to $a_1$, we see the characteristic
values of effects caused by the interference is about 35-50\%.

At large recoils as in $B_c^+\to \psi \pi^+(\rho^+)$, the
spectator picture of transition can be broken by the hard gluon
exchanges \cite{Gers}. The spin effects in such decays were
studied in \cite{Pakh}. However, we emphasize that the significant
rates of $B_c$ decays to the P- and D-wave charmonium states point
out that the corrections in the second order of the heavy-quark
velocity in the heavy quarkonia under study could be quite
essential and they can suppress the corresponding decay rates,
since the relative momentum of heavy quarks inside the quarkonium
if different from zero should enhance the virtuality of gluon
exchange, which suppresses the decay amplitudes.

For the widths of non-leptonic $c$-quark decays we have found that
the sum rule estimates are greater than those of potential
models\footnote{See also recent discussions of the $B_c$ decays in
\cite{verma,giri,Nobes,Mannel,Ma,Ivanov2,Faust'}.}. In this
respect we check that the QCD SR calculations are consistent with
the inclusive ones. So, we sum up the calculated exclusive widths
and estimate the total width of $B_c$ meson as shown in Fig.
\ref{life}, which points to a good agreement of our calculations
with those of OPE and semi-inclusive estimates.

Another interesting point is the possibility to extract the
factorization parameters $a_1$ and $a_2$ in the $c$-quark decays
by measuring the branching ratios
\begin{eqnarray}
\frac{\Gamma[B_c^+\to B^+\bar K^0]}{\Gamma[B_c^+\to B^0 K^+]} &=&
\frac{\Gamma[B_c^+\to B^+\bar K^{*0}]}{\Gamma[B_c^+\to B^0
K^{*+}]} = 
\frac{\Gamma[B_c^+\to B^{*+}\bar K^0]}{\Gamma[B_c^+\to B^{*0}
K^+]} = \frac{\Gamma[B_c^+\to B^{*+}\bar K^{*0}]}{\Gamma[B_c^+\to
B^{*0} K^{*+}]} =
\label{fact} 
\frac{\Gamma_0}{\Gamma_+}=
\left|\frac{V_{cs}}{V_{cd}^2}\right|^2\,\left(\frac{a_2}{a_1}\right)^2.
\end{eqnarray}
This procedure can give the test for the factorization approach
itself.

The suppressed decays caused by the flavor changing neutral
currents were studied in \cite{rare}.

The {\sf CP}-violation in the $B_c$ decays can be investigated in
the same manner as made in the $B$ decays. The expected {\sf
CP}-asymmetry of ${\cal A}(B_c^\pm \to J/\psi D^\pm)$ is about
$4\cdot 10^{-3}$, when the corresponding branching ratio is
suppressed as $10^{-4}$ \cite{CPBc}. Thus, the direct study of
{\sf CP}-violation in the $B_c$ decays is practically difficult
because of low relative yield of $B_c$ with respect to ordinary
$B$ mesons: $\sigma(B_c)/\sigma(B) \sim 10^{-3}$. A
model-independent way to extract the CKM angle $\gamma$ based on
the measurement of two reference triangles was independently
offered by Masetti, Fleischer and Wyler in \cite{CPBc} by
investigating the modes with the neutral charmed meson in the
final state (see the corresponding subsection below).

Another possibility is the lepton tagging of $B_s$ in the
$B_c^\pm\to B_s^{(*)} l^\pm \nu$ decays for the study of mixing
and {\sf CP}-violation in the $B_s$ sector \cite{Quigg}.

\subsubsection{{CP}-Violation in Decays of
$B_{c}$ Meson}

The $B_c$ meson first observed by the CDF collaboration at FNAL
\cite{cdf} is expected to be copiously produced in the future
experiments at hadron colliders \cite{revbc} with facilities
oriented to the study of fine effects in the heavy quark
interactions such as the parameters of CP-violation and charged
weak current mixing\footnote{See, for instance, the program on the
B physics at Tevatron \cite{BRunII}.}. So, one could investigate
the spectroscopy, production mechanism and decay features of $B_c$
\cite{Bj,revbc,OPEBc,KT,KLO,KKL,exBc} with the incoming sample of
several billion events. In such circumstances, in addition to the
current success in the experimental study of decays with the
CP-violation in the gold-plated mode of neutral $B$-meson by the
BaBar and Belle collaborations \cite{BB} allowing one to extract
the CKM-matrix angle $\beta$ in the unitarity triangle, a possible
challenge is whether one could get an opportunity to extract some
information about the CKM unitarity triangle from the $B_c$
physics in a model independent way or not. The theoretical
principal answer is one can do it. Indeed, there is an intriguing
opportunity to extract the angle $\gamma$ in the model-independent
way using the strategy of reference triangles \cite{Gronau} in the
decays of doubly heavy hadrons. This ideology for the study of
CP-violation in $B_c$ decays was originally offered by M.Masetti
\cite{CPBc}, independently investigated by R.Fleischer and D.Wyler
\cite{CPBc} and extended to the case of doubly heavy
baryons\footnote{A review on the physics of doubly heavy baryons
is given in \cite{QQq}.} in \cite{CPQQq}.

Let us point out necessary conditions to extract the CP-violation
effects in the model-independent way.
\begin{enumerate}
\item Interference. The measured quantities have to involve the
amplitudes including both the CP-odd and CP-even phases. \item
  Exclusive channels. The hadronic final state has to be fixed in
  order to isolate a definite flavor contents and, hence, the
  definite matrix elements of CKM matrix, which can exclude the
  interference of two CP-odd phases with indefinite CP-even phases due
  to strong interactions at both levels of the quark structure and the
  interactions in the final state.
\item Oscillations. The definite involvement of the CP-even phase
is
  ensured by the oscillations taking place in the systems of
  neutral $B$ or $D$ mesons, wherein the CP-breaking effects can be
  systematically implemented.
\item Tagging. Once the oscillations are involved, the tagging of
both
  the flavor and CP eigenstates is necessary for the complete
  procedure.
\end{enumerate}
The gold-plated modes in the decays of neutral $B$ mesons involve
the oscillations of mesons themselves and, hence, they require the
time-dependent measurements. In contrast, the decays of doubly
heavy hadrons such as the $B_c$ meson and $\Xi_{bc}$ baryons with
the neutral $D^0$ or $\bar D^0$ meson in the final state do not
require the time-dependent measurements. The triangle ideology is
based on the direct determination of absolute values for the set
of four decays, at least: the decays of hadron in the tagged $D^0$
meson, the tagged $\bar D^0$ meson, the tagged CP-even
state\footnote{The CP-odd states of $D^0$ can be used, too.
However, their registration requires the detection of CP-even
state of $K^0$, which can be complicated because of a detector
construction, say, by a long base of $K^0$ decay beyond a tracking
system.} of $D^0$, and the decay of the anti-hadron into the
tagged CP-even state of $D^0$. To illustrate, let us consider the
decays of
$$
B^+_c\to D^0 D_s^+, \quad\mbox{and}\quad B^+_c\to \bar D^0 D_s^+.
$$

\begin{figure*}[th]
  \begin{center}
\setlength{\unitlength}{1.3mm} \hspace*{2cm}
\begin{picture}(90,25)
\put(-10,0){\epsfxsize=55\unitlength \epsfbox{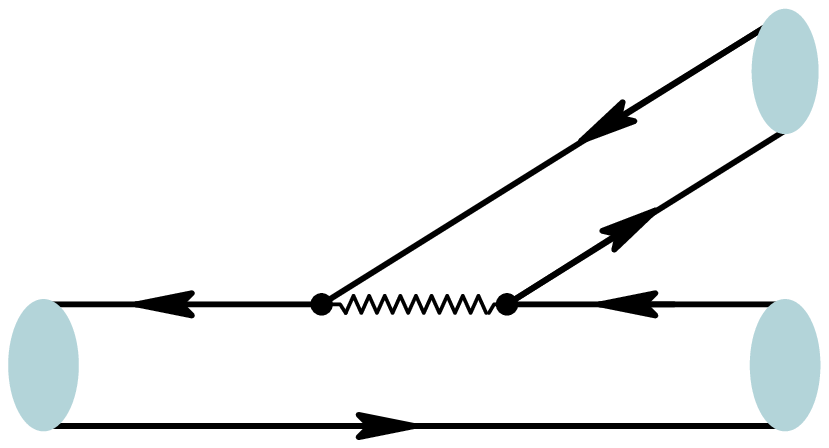}}
\put(33,0){\epsfxsize=55\unitlength \epsfbox{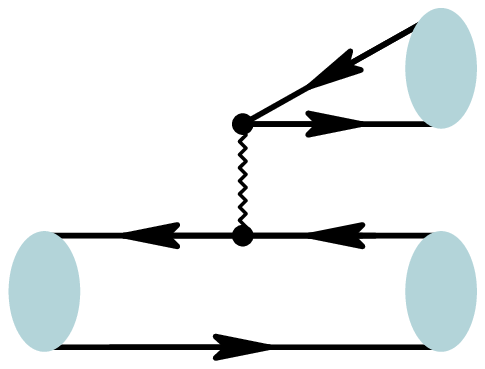}}
\put(-2.3,6.5){$B^+_{c}$} \put(34.,6.5){$D^+_{s}$}
\put(34.2,21){$D^0$} \put(47.2,6.55){$B^+_{c}$}
\put(67.5,6.6){$D^0$} \put(67.5,18.1){$D^+_{s}$}
\put(5.,11.5){$\scriptstyle b$} \put(15.,2.){$\scriptstyle c$}
\put(26.,20.8){$\scriptstyle u$} \put(27,15.){$\scriptstyle c$}
\put(29,11.5){$\scriptstyle s$} \put(53.5,12){$\scriptstyle b$}
\put(63.5,11.5){$\scriptstyle u$} \put(57.5,2){$\scriptstyle c$}
\put(63.2,20.2){$\scriptstyle s$} \put(63,14.2){$\scriptstyle c$}
\end{picture}
    \caption{The diagrams of $\bar b$-quark decay contributing to the weak
     transition $B^+_c\to D^0 D_s^+$.}
    \label{fig:1}
  \end{center}
\end{figure*}
\begin{figure*}[th]
  \begin{center}
\setlength{\unitlength}{1.3mm} \hspace*{0cm}
\begin{picture}(90,25)
\put(-10,2){\epsfxsize=55\unitlength \epsfbox{1g.eps}}
\put(42,1){\epsfxsize=55\unitlength \epsfbox{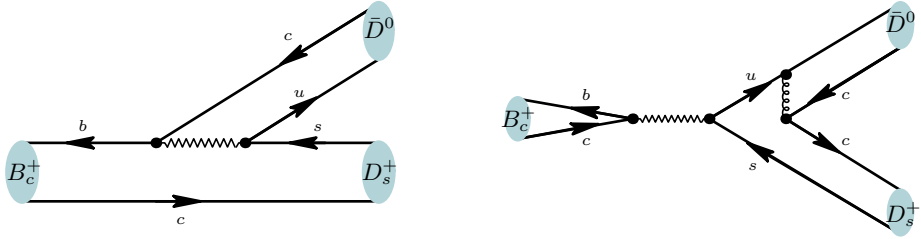}}
\put(-2.3,8.5){$B^+_{c}$} \put(34.,8.5){$D^+_{s}$}
\put(34.2,23){$\bar D^0$} \put(48.2,13.9){$B^+_{c}$}
\put(87.5,4.2){$D^+_{s}$} \put(87.5,23.6){$\bar D^0$}
\put(5.,13.5){$\scriptstyle b$} \put(15.,4.){$\scriptstyle c$}
\put(26.,22.8){$\scriptstyle c$} \put(27,17.){$\scriptstyle u$}
\put(29,13.5){$\scriptstyle s$} \put(56.5,16.7){$\scriptstyle b$}
\put(73.5,9.5){$\scriptstyle s$} \put(56.5,12){$\scriptstyle c$}
\put(73.2,18.7){$\scriptstyle u$} \put(83,16.8){$\scriptstyle c$}
\put(83,12.){$\scriptstyle c$}
\end{picture}
    \caption{The diagrams of $\bar b$-quark decay contributing to the weak
     transition $B^+_c\to \bar D^0 D_s^+$.}
    \label{fig:1a}
  \end{center}
\end{figure*}

The corresponding diagrams with the decay of $\bar b$-quark are
shown in Figs. \ref{fig:1} and \ref{fig:1a}. We stress that two
diagrams of the decay to $D^0$ have the additional negative sign
caused by the Pauli interference of two charmed quarks, which,
however, completely compensated after the Fierz transformation for
the corresponding Dirac matrices.

The exclusive modes make the penguin terms to be excluded, since
the penguins add an even number of charmed quarks, i.e. two or
zero, while the final state contains two charmed quarks including
one from the $\bar b$ decay and one from the initial state.
However, the diagram with the weak annihilation of two
constituents, i.e. the charmed quark and beauty anti-quark in the
$B_c^+$ meson, can contribute in the next order in $\alpha_s$ as
shown in Fig. \ref{fig:1a} for the given final state.
Nevertheless, we see that such the diagrams have the same
weak-interaction structure as at the tree level. Therefore, they
do not break the consideration under interest. The magnitude of
$\alpha_s$-correction to the absolute values of corresponding
decay widths is discussed in \cite{CPBcKis}.

\begin{figure}[th]
  \begin{center}
\setlength{\unitlength}{1mm}
\begin{picture}(80,50)
\put(10,7){\epsfxsize=75\unitlength \epsfbox{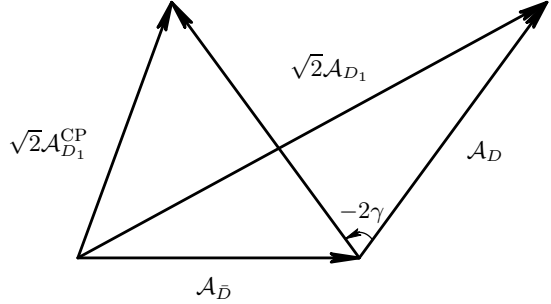}}
\put(32,4){${\cal A}_{\bar D}$} \put(68,22){${\cal A}_{D}$}
\put(51.,14){$-2\gamma$} \put(44.5,33){$\sqrt{2}{\cal A}_{D_1}$}
\put(7,23){$\sqrt{2}{\cal A}_{D_1}^{\mbox{\sc cp}}$}
\end{picture}
    \caption{The reference-triangles.}
    \label{fig:3}
  \end{center}
\end{figure}

Thus, the CP-odd phases of decays under consideration are
determined by the tree-level diagrams shown in Figs. \ref{fig:1}
and \ref{fig:1a}. Therefore, we can write down the amplitudes in
the following form:
\begin{equation}
  \label{eq:3}
  {\cal A}(B^+_c\to D^0 D_s^+) \stackrel{\mbox{\tiny def}}{=}
  {\cal A}_{D} = V^*_{ub} V_{cs}\cdot {\cal M}_{D},\qquad
  {\cal A}(B^+_c\to \bar D^0 D_s^+) \stackrel{\mbox{\tiny def}}{=}
  {\cal A}_{\bar D} = V^*_{cb} V_{us}\cdot {\cal M}_{\bar D},
\end{equation}
where ${\cal M}_{\bar D,\,D}$ denote the CP-even factors depending
on the dynamics of strong interactions. Using the definition of
angle $\gamma$
$$
\gamma \stackrel{\mbox{\tiny def}}{=} -{\rm arg}\left[\frac{V_{ub}
V^*_{cs}} {V_{cb} V^*_{us}}\right],
$$
for the CP-conjugated channels\footnote{For the sake of simplicity
  we put the overall phase of arg $V_{cb} V^*_{us}=0$, which
  corresponds to fixing the representation of the CKM matrix, e.g. by
  the Wolfenstein form \cite{Wolf}.} we find
\begin{equation}
  \label{eq:4}
  {\cal A}(B^-_c\to \bar D^0 D_s^-)= e^{-2{\rm i}\gamma}
  {\cal A}_{D}, \qquad
  {\cal A}(B^-_c\to D^0 D_s^-)=  {\cal A}_{\bar D}.
\end{equation}
We see that the corresponding widths for the decays to the flavor
tagged modes coincide with the CP-conjugated ones. However, the
story can be continued by using the definition of CP-eigenstates
for the oscillating $D^0\leftrightarrow \bar D^0$
system\footnote{The
  suppressed effects of CP-violation in the oscillations of neutral
  $D$ mesons are irrelevant here, and we can neglect them in the sound
  way.},
$$
D_{1,\,2} =\frac{1}{\sqrt{2}}(D^0\pm \bar D^0),
$$
so that we straightforwardly get
\begin{eqnarray}
  \label{eq:5}
 \sqrt{2}{\cal A}(B^+_c\to D_s^+ D_1)
 \stackrel{\mbox{\tiny def}}{=}
 \sqrt{2} {\cal A}_{D_1} &=& {\cal A}_{D}+{\cal A}_{\bar D},\\[2mm]
 \sqrt{2}{\cal A}(B^-_c\to D_s^- D_1)
 \stackrel{\mbox{\tiny def}}{=}
 \sqrt{2} {\cal A}_{D_1}^{\mbox{\sc cp}}& = &e^{-2{\rm i}\gamma}
 {\cal A}_{D}+{\cal A}_{\bar D}.
\label{eq:6}
\end{eqnarray}
The complex numbers entering (\ref{eq:5}) and (\ref{eq:6})
establish two triangles with the definite angle $2\gamma$ between
the vertex positions as shown in Fig. \ref{fig:3}.
Thus, due to the unitarity, the measurement of four absolute
values
\begin{eqnarray}
  \label{eq:7}
  |{\cal A}_{D}| = |{\cal A}(B^+_c\to D_s^+ D^0)|,
&\quad &
  |{\cal A}_{\bar D}| = |{\cal A}(B^+_c\to D_s^+ \bar D^0)|, \nonumber\\[2mm]
  |{\cal A}_{D_1}| = |{\cal A}(B^+_c\to D_s^+ D_1)|,
&\quad &
  |{\cal A}_{D_1}^{\mbox{\sc cp}}| = |{\cal A}(B^-_c\to D_s^- D_1)|,
\end{eqnarray}
can constructively reproduce the angle $\gamma$ in the
model-independent way.

The above triangle-ideology can be implemented for the analogous
decays to the excited states of charmed mesons in the final state.

The residual theoretical challenge is to evaluate the
characteristic widths or branching fractions. We address this
problem and analyze the color structure of amplitudes. So, we find
that the matrix elements under interest have the different
magnitudes of color suppression, so that at the tree level we get
${\cal A}_{D}\sim O(\sqrt{N_c})$ and ${\cal A}_{\bar D}\sim
O(1/\sqrt{N_c})$, while the ratio of relevant CKM-matrix elements,
$$
\left|\frac{V_{ub} V^*_{cs}}{V_{cb} V^*_{us}}\right|\sim O(1)
$$
with respect to the small parameter of Cabibbo angle, $\lambda =
\sin\theta_C$, which one can easily find in the Wolfenstein
parametrization
$$
V_{\mbox{\sc ckm}} = \left(\begin{array}{ccc}
V_{ud} & V_{us} & V_{ub} \\[2mm]
V_{cd} & V_{cs} & V_{cb} \\[2mm]
V_{td} & V_{ts} & V_{tb}
\end{array}\right)
= \left(\begin{array}{ccc}
1-\frac{1}{2}\lambda^2 & \lambda & A \lambda^3 (\rho-{\rm i}\eta) \\[2mm]
-\lambda & 1-\frac{1}{2}\lambda^2  & A \lambda^2 \\[2mm]
A \lambda^3 (1-\rho-{\rm i}\eta) & -A \lambda^2 & 1
\end{array}\right).
$$
Nevertheless, the interference of two diagrams in the decays of
$B_c^+$ to the $D^0$ meson is destructive, and the absolute values
of the amplitudes ${\cal A}_{D}$ and ${\cal A}_{\bar D}$ become
close to each other. Thus, we expect that the sides of the
reference-triangles are of the same order of magnitude, which
makes the method to be an attractive way to extract the
\mbox{angle $\gamma$}.

The predictions of QCD sum rules for the exclusive decays of $B_c$
are summarized in Table \ref{common2} at the fixed values of
factors $a_{1,2}$ and lifetime. For the sake of completeness and
comparison we show the estimates for the channels with the neutral
$D$ meson and charged one $D^+$ as well as for the vector states
in addition to the pseudoscalar ones.

First, we see that the similar decay modes without the strange
quark in the final state can be, in principle, used for the same
extraction of CKM angle $\gamma$, however, this channels are more
problematic from the methodic point of view, since the sides of
reference-triangles significantly differ from each
other\footnote{The ratio of widths is basically determined by the
factor of $|V_{cb} V_{ud} a_2|^2/|V_{ub} V_{cd} a_1|^2\sim 110$,
if we ignore the interference effects.}, so that the measurements
have to be extremely accurate in order to get valuable information
on the angle. Indeed, we should accumulate a huge statistics for
the dominant mode in order to draw any conclusion on the
consistency of triangle with a small side.

Second, the decay modes with the vector neutral $D$ meson in the
final state are useless for the purpose of the CKM measurement
under the approach discussed. However, the modes with the vector
charged $D^*$ and $D_s^*$ mesons can be important for the
procedure of $\gamma$ extraction. This note could be essential for
the mode with $D^{*+}\to D^0 \pi^+$ and $D^0\to K^-\pi^+$, but, in
this case, the presence of neutral charmed meson should be
carefully treated in order to avoid the misidentification with the
primary neutral charmed meson. In other case, we should use the
mode with the neutral pion $D^{*+}\to D^+ \pi^0$, which detection
in an experimental facility could be problematic. The same note is
applicable for the vector $D_s^{*+}$ meson, which radiative
electromagnetic decay is problematic for the detection, too, since
the photon could be loosed. However, the lose of the photon for
the fully reconstructed $D_s^+$ and $B_c^+$ does not disturb the
analysis.

\begin{table*}[th]
\caption{Branching ratios of exclusive $B_c^+$ decays at the fixed
choice of factors: $a_1^b=1.14$ and $a_2^b=-0.20$ in the
non-leptonic decays of $\bar b$ quark. The lifetime of $B_c$ is
appropriately normalized by $\tau[B_c] \approx 0.45$ ps.}
\label{common2}
\begin{center}
\begin{tabular}{|l|rr|}
\hline
~~~~~Mode & \multicolumn{2}{c|}{BR, $10^{-6}$}\\
\hline
 $B_c^+ \rightarrow D^+
\overline D^{\hspace{1pt}\raisebox{-1pt}{$\scriptscriptstyle 0$}}$
 & 53 & [18]\\
 $B_c^+ \rightarrow D^+
\overline D^{\hspace{1pt}\raisebox{-1pt}{$\scriptscriptstyle
*0$}}$
 & 75 & [20]\\
 $B_c^+ \rightarrow  D^{\scriptscriptstyle *+}
\overline D^{\hspace{1pt}\raisebox{-1pt}{$\scriptscriptstyle 0$}}$
 & 49 & [9]\\
 $B_c^+ \rightarrow  D^{\scriptscriptstyle *+}
\overline D^{\hspace{1pt}\raisebox{-1pt}{$\scriptscriptstyle
*0$}}$
 & 330 & [120]\\
 $B_c^+ \rightarrow D_s^+ \overline
D^{\hspace{1pt}\raisebox{-1pt}{$\scriptscriptstyle 0$}}$
 & 4.8 & [1]\\
 $B_c^+ \rightarrow D_s^+
\overline D^{\hspace{1pt}\raisebox{-1pt}{$\scriptscriptstyle
*0$}}$
 & 7.1 & [1.2]\\
 $B_c^+ \rightarrow  D_s^{\scriptscriptstyle *+} \overline
D^{\hspace{1pt}\raisebox{-1pt}{$\scriptscriptstyle 0$}}$
 & 4.5 & [0.5]\\
 $B_c^+ \rightarrow  D_s^{\scriptscriptstyle *+}
\overline D^{\hspace{1pt}\raisebox{-1pt}{$\scriptscriptstyle
*0$}}$
 & 26 & [2]\\
\hline
\end{tabular}
\begin{tabular}{|l|rr|}
\hline
~~~~~Mode & \multicolumn{2}{c|}{BR, $10^{-6}$}\\
\hline
 $B_c^+ \rightarrow D^+
D^{\hspace{1pt}\raisebox{-1pt}{$\scriptscriptstyle 0$}}$
 & 0.32 & [0.1]\\
 $B_c^+ \rightarrow D^+
D^{\hspace{1pt}\raisebox{-1pt}{$\scriptscriptstyle *0$}}$
 & 0.28 & [0.07]\\
 $B_c^+ \rightarrow  D^{\scriptscriptstyle *+}
D^{\hspace{1pt}\raisebox{-1pt}{$\scriptscriptstyle 0$}}$
 & 0.40 & [0.4]\\
 $B_c^+ \rightarrow  D^{\scriptscriptstyle *+}
D^{\hspace{1pt}\raisebox{-1pt}{$\scriptscriptstyle *0$}}$
 & 1.59 & [0.4]\\
 $B_c^+ \rightarrow D_s^+
D^{\hspace{1pt}\raisebox{-1pt}{$\scriptscriptstyle 0$}}$
 & 6.6 & [1.7]\\
 $B_c^+ \rightarrow D_s^+
D^{\hspace{1pt}\raisebox{-1pt}{$\scriptscriptstyle *0$}}$
 & 6.3 & [1.3]\\
 $B_c^+ \rightarrow  D_s^{\scriptscriptstyle *+}
D^{\hspace{1pt}\raisebox{-1pt}{$\scriptscriptstyle 0$}}$
 & 8.5 & [8.1]\\
 $B_c^+ \rightarrow  D_s^{\scriptscriptstyle *+}
D^{\hspace{1pt}\raisebox{-1pt}{$\scriptscriptstyle *0$}}$
 & 40.4 & [6.2]\\
\hline
\end{tabular}
\end{center}
\end{table*}

In the BTeV \cite{BRunII} and LHCb \cite{LHCB} experiments one
expects the $B_c$ production at the level of several billion
events. Therefore, we predict $10^4-10^5$ decays of $B_c$ in the
gold-plated modes under interest. The experimental challenge is
the efficiency of detection. One usually get a 10\%-efficiency for
the observation of distinct secondary vertices outstanding from
the primary vertex of beam interaction. Next, we have to take into
account the branching ratios of $D_s$ and $D^0$ mesons. This
efficiency crucially depends on whether we can detect the neutral
kaons and pions or not. So, for the $D_s$ meson the corresponding
branching ratios grow from 4\% (no neutral $K$ and $\pi$) to 25\%.
The same interval for the neutral $D^0$ is from 11 to 31\%. The
detection of neutral kaon is necessary for the measurement of
decay modes into the CP-odd state $D_2$ of the neutral $D^0$
meson, however, we can omit this cross-check channel from the
analysis dealing with the CP-even state of $D_1$. The
corresponding intervals of branching ratios reachable by the
experiment are from 0.5 to 1.3\% for the CP-even state and from
1.5 to 3.8\% for the CP-odd state of $D^0$. The pessimistic
estimate for the product of branching ratios is about $2\cdot
10^{-4}$, which results in $2-20$ reconstructed events. Thus, an
acceptance of experimental facility and an opportunity to detect
neutral pions and kaons as well as reliable estimates of total
cross section for the $B_c$ production in hadronic collisions are
of importance in order to make expectations more accurate.

\section{Conclusion}

We have reviewed the current status of theoretical predictions for
the decays of $B_c$ meson.

We have found that the various approaches: OPE, Potential models
and QCD sum rules, result in the close estimates, while the SR as
explored for the various heavy quark systems, lead to a smaller
uncertainty due to quite an accurate knowledge of the heavy quark
masses. So, summarizing we expect that the dominant contribution
to the $B_c$ lifetime is given by the charmed quark decays ($\sim
70\%$), while the $b$-quark decays and the weak annihilation add
about 20\% and 10\%, respectively. The predictions have been
presented in the form of long tables with the numerical values of
branching ratios. There are many physical points, which can be
investigated in the $B_c$ decays. We hope that this possibility
could be, in various aspects, studied in the forthcoming
experimental runs at hadron colliders, where the production rate
about several billions $B_c$ actually opens such the opportunity.

The author thanks many members of the community of physicists
working in the field of heavy quarkonium for discussions and
stimulating results concerning for the $B_c$ meson.

This work is partially supported by the grants of RFBR
01-02-99315, 01-02-16585, the grants of RF president for young
Doctors of Science MD-297.2003.02 and for scientific schools
NSc-1303.2003.2.

\end{document}